\documentclass[12pt]{article}

\pdfoutput=1

\usepackage{cite}

\usepackage{amssymb,latexsym}

\usepackage{epstopdf}

\usepackage{graphics,epsfig}

\usepackage{color}

\usepackage{mathrsfs}

\DeclareMathAlphabet{\mathpzc}{OT1}{pzc}{m}{it}

\let\a=\alpha \let\b=\beta \let\g=\gamma \let\d=\delta \let\e=\epsilon
\let\z=\zeta  \let\th=\theta  \let\k=\kappa
\let\l=\lambda \let\m=\mu \let\n=\nu \let\x=\xi \let\p=\pi %\let\r=\rho
\let\s=\sigma   \let\f=\phi  
 
      \let\G=\Gamma \let\D=\Delta \let\Th=\Theta 
\let\X=\Xi  \let\S=\Sigma  \let\Y=\Psi
 
\let\la=\label  
  
\def\nn{\nonumber} \def\bd{\begin{document}} \def\ed{\end{document}}
\def\ds{\documentstyle} \let\fr=\frac \let\bl=\bigl \let\br=\bigr
\let\Br=\Bigr \let\Bl=\Bigl
\let\bm=\bibitem
\let\na=\nabla
\def\tU{{\widetilde U}}
\let\pa=\partial \let\ov=\overline
\def\ie{{\it i.e.\ }}
\newcommand{\be}{\begin{equation}}
\newcommand{\ee}{\end{equation}}
\def\ba{\begin{array}}
\def\ea{\end{array}}
\def\ft#1#2{{\textstyle{{\scriptstyle #1}\over {\scriptstyle #2}}}}
\def\fft#1#2{{#1 \over #2}}
\def\F#1#2{{ F_{#1}^{(#2)} }}
\def\cF#1#2{{ {\cal F}_{#1}^{(#2)} }}

\def\R{{\bf R}}
\def\sst#1{{\scriptscriptstyle #1}}
\def\oneone{\rlap 1\mkern4mu{\rm l}}
\def\e7{E_{7(+7)}}
\def\td{\tilde}
\def\wtd{\widetilde}
\def\im{{\rm i}}
\def\bog{Bogomol'nyi\ }
\newcommand{\ho}[1]{$\, ^{#1}$}
\newcommand{\hoch}[1]{$\, ^{#1}$}
\newcommand{\bea}{\begin{eqnarray}}
\newcommand{\eea}{\end{eqnarray}}
\newcommand{\ra}{\rightarrow}
\newcommand{\lra}{\longrightarrow}
\newcommand{\Lra}{\Leftrightarrow}
\newcommand{\ap}{\alpha^\prime}
\newcommand{\bp}{\tilde \beta^\prime}
\newcommand{\cB}{{\cal B}}
\newcommand{\cO}{{\cal O}}
\newcommand{\vecx}{\vec{x}}
\newcommand{\vecy}{\vec{y}}
\newcommand{\vecp}{\vec{p}}
\newcommand{\vecq}{\vec{q}}
\newcommand{\tr}{{\rm tr} }
\newcommand{\Tr}{{\rm Tr} }
\newcommand{\NP}{Nucl. Phys. }

\newcommand{\cL}{{\cal L}}
\newcommand{\cA}{{\cal A}}
\newcommand{\cT}{{\cal T}}
\newcommand{\cD}{{\cal D}}
\newcommand{\cH}{{\cal H}}

\def\sst#1{{\scriptscriptstyle #1}}
\def\0{{\sst{(0)}}}
\def\1{{\sst{(1)}}}
\def\2{{\sst{(2)}}}
\def\3{{\sst{(3)}}}
\def\4{{\sst{(4)}}}
\def\5{{\sst{(5)}}}
\def\6{{\sst{(6)}}}
\def\7{{\sst{(7)}}}
\def\8{{\sst{(8)}}}
\def\9{{\sst{(9)}}}
\def\p{{\sst{(p)}}}
\def\q{{\sst{(q)}}}
\def\ve{\varepsilon}
\def\vf{\varphi}
\def\F{\Phi}
\def\wg{\wedge}

\def\thb{\bar{\theta}}
\def\Thb{\bar{\Theta}}
\def\barp{\bar{p}}
\def\barq{\bar{q}}
\def\barc{\bar{c}}
\def\bard{\bar{d}}
\def\e{\epsilon}

\def\Cb{\bar{C}}

\def \bi{\bibitem}
\def \la {\label}

\def \l {\lambda}
\def\foot{\footnote}
\def \tl  {{\tilde \l}}
\def \sql {{\sqrt \l}}
\def \adss {$AdS_5 \times S^5$\ }
\newcommand{\rf}[1]{(\ref{#1})}
\def \ov {\over}

\def\th{\theta}
\def\Th{\Theta}
\def\vth{\vartheta}
\def\btheta{{\bar\theta}}
\def\ttheta{{{\tilde\theta}}}
\def\bttheta{{{\bar\ttheta}}}
\def\vth{\vartheta}

\def\ra{\rightarrow}
\def\N{\nabla}
\def\F{{\cal F}}
\def\uM{\underline{M}}
\def\uA{\underline{A}}
\def\uN{\underline{N}}
\def\uP{\underline{P}}
\def\ua{\underline{a}}
\def\ub{\underline{b}}
\def\uc{\underline{c}}
\def\ud{\underline{d}}
\def\ue{\underline{e}}
\def\uf{\underline{f}}
\def\ui{\underline{i}}
\def\uj{\underline{j}}
\def\uk{\underline{k}}
\def\ul{\underline{l}}
\def\ual{\underline{\alpha}}
\def\ube{\underline{\beta}}
\def\um{\underline{m}}
\def\un{\underline{n}}
\def\up{\underline{p}}
\def\uq{\underline{q}}
\def\ur{\underline{r}}
\def\us{\underline{s}}
\def\umu{\underline{\mu}}
\def\unu{\underline{\nu}}
\def\ula{\underline{\l}}
\def\uka{\underline{\k}}
\def\usi{\underline{\s}}
\def\urh{\underline{\r}}
\def\cc{\circ}
\def\eqv{\equiv}

\def\ni{\noindent}

\def\Ep{E^{{}^{(+)}}}
\def\Em{E^{{}^{(-)}}}

\def\Mp{M^{{}^{(+)}}}
\def\Mm{M^{{}^{(-)}}}

\def \ha{{1\ov 2}}

\def\r{\rho}

\def\Y{{\rm Y}}
\def\X{{\rm X}}
\def\tY{\tilde{\rm Y}}
\def\tX{\tilde{\rm X}}
\def\dY{\dot{\rm Y}}
\def\dX{\dot{\rm X}}

\def \J {\mathcal{J}}
\def \del {\partial}

\def\dF{\dot{F}}
\def\dG{\dot{G}}
\def\df{\dot{f}}
\def \E {{\cal E}}
\def \S {{\cal S}}
\def \J {{\cal J}}

\def\ms{\mathcal{S}}
\def\mj{\mathcal{J}}
\def\soj{\fr{\ms}{\mj}}
\def \R {{\bf R}}
\def \om {\omega}
\def \bE {\bar E}
\def \x {{\cal X}}

\def \bi{\bibitem}
\def \la {\label}

\def \l {\lambda}
\def\foot{\footnote}
\def \tl  {{\tilde \l}}
\def \sql {{\sqrt \l}}
\def \adss {$AdS_5 \times S^5$\ }
\def \ov {\over}

\def \varpi {{\rm w}}

\def\thb{\bar{\theta}}
\def\Thb{\bar{\Theta}}
\def\mb{\bar{\m}}
\def\ab{\bar{\a}}
\def\zb{\bar{z}}
\def\psib{\bar{\psi}}
\def\barp{\bar{p}}
\def\barq{\bar{q}}
\def\barc{\bar{c}}
\def\bard{\bar{d}}
\def\e{\epsilon}
\def\wb{\bar{w}}
\def\lb{\bar{\l}}
\def\Jb{\bar{J}}
\def\Nb{\bar{N}}
\def\Zb{\bar{Z}}
\def\pab{\bar{\pa}}

\def\At{\tilde{A}}
\def\Bt{\tilde{B}}
\def\Ct{\tilde{C}}
\def\Dt{\tilde{D}}
\def\Et{\tilde{E}}
\def\Ft{\tilde{F}}
\def\Gt{\tilde{G}}
\def\Ht{\tilde{H}}
\def\Kt{\tilde{K}}
\def\Mt{\tilde{M}}
\def\Nt{\tilde{N}}
\def\Rt{\tilde{R}}
\def\at{\tilde{a}}
\def\bt{\tilde{b}}
\def\ct{\tilde{c}}
\def\dt{\tilde{d}}
\def\et{\tilde{e}}
\def\ft{\tilde{f}}
\def\htil{\tilde{h}}
\def\gt{\tilde{g}}
\def\nt{\tilde{n}}
\def\mut{\tilde{\mu}}
\def\nut{\tilde{\nu}}
\def\pht{\tilde{\f}}
\def\vft{\tilde{\vf}}

\def\rht{\tilde{\rho}}

\def\asth{\hat{*}}
\def\phh{\hat{\phi}}

\def\bA{{\bf A}}

\def\ola{\overleftarrow}
\def\ora{\overrightarrow}
\def\alt{\tilde{\a}}

\def\eh{\hat{e}}
\def\eph{\hat{\e}}
\def\ph{\hat{p}}
\def\alh{\hat{\a}}
\def\beh{\hat{\b}}
\def\gah{\hat{\g}}
\def\Fh{\hat{F}}
\def\muh{\hat{\m}}
\def\nuh{\hat{\n}}
\def\thh{\hat{\th}}
\def\rhh{\hat{\r}}
\def\dh{\hat{d}}
\def\ih{\hat{i}}
\def\jh{\hat{j}}
\def\hh{\hat{h}}
\def\nh{\hat{n}}
\def\gh{\hat{g}}
\def\kh{\hat{k}}
\def\deh{\hat{\d}}
\def\wh{\hat{w}}
\def\lah{\hat{\l}}
\def\Ah{\hat{A}}
\def\Kh{\hat{K}}
\def\Nh{\hat{N}}
\def\Rh{\hat{R}}
\def\Ch{\hat{C}}
\def\Omh{\hat{\Omega}}

\def\xh{\hat{x}}

\def\ps{\rlap{\, /}\;\,p }
\def\ks{\rlap{\, /}\;\,k }

\def\gym{g_{YM}}

\def\adot{\dot{a}}
\def\bdot{\dot{b}}
\def\bpa{\bar{\pa}}

\def\pr{\prime}
\def\ssk{\medskip}
\def\clb{\color{blue}}
\def\clr{\color{red}}
\def\clg{\color{green}}

\def\bfA{{\bf A}}
\def\bfB{{\bf B}}
\def\bfK{{\bf K}}
\def\bfU{{\bf U}}
\def\bfX{{\bf X}}
\def\bfY{{\bf Y}}
\def\bfZ{{\bf Z}}
\def\bfg{{\bf g}}
\def\bfn{{\bf n}}

\begin{document}

\overfullrule=0pt
\parskip=2pt
\parindent=12pt
\headheight=0in \headsep=0in \topmargin=0in
\oddsidemargin=0in

\vspace{ -3cm}
\thispagestyle{empty}
%\vspace{1cm}
%\begin{flushright}
%Preprint DFPD 01/TH/\\
%hep-th/
%\end{flushright}

 \vspace{0.1cm}

\setcounter{equation}{0}
\setcounter{footnote}{0}
\setcounter{section}{0}

\begin{center}

{\Large\bf On 4D covariance of Feynman diagrams of Einstein gravity}

\vskip 0.8cm

 \vspace{.5cm}

\vspace{0.5cm}
I. Y. Park
\\

\vspace{0.3cm}

%{\it Center for Quantum Spacetime, Sogang University\\
%Shinsu-dong 1, Mapo-gu, 121-742 South Korea \\
%}

%{\it Department of Physics, Hanyang University \\
%Seoul 133-791, Korea}\\

\vspace{0.3cm}
{\it Department of Applied Mathematics,
Philander Smith College %\footnote{Home institute}
                               \\
Little Rock, AR 72223, USA \\
inyongpark05@gmail.com
}

\end{center}

 \vspace{0.1cm}

\begin{abstract}

It has been observed in \cite{Park:2014tia} that the physical states of the ADM formulation of 4D Einstein gravity holographically reduce and can be described by a 3D language. The main theme of the present work is two 4D covariance issues.
 A proper handling of the trace piece of the fluctuation metric through gauge-fixing is the key to one of the covariance issues. Although the unphysical character of the trace piece has been long known, it has not been taken care of in a manner suitable for the Feynman diagram computations. As for the second covariance issue, a renormalization program can be carried out covariantly to any loop order at intermediate steps, thereby maintaining the 4D covariance; it is only at the final stage that one should consider the 3D physical external states. With the physical external states, the 1PI effective action reduces to 3D and renormalizability is restored just as in the entirely-3D approach of \cite{Park:2014noa}.
Paying a careful attention to the trace piece of the fluctuation metric we revisit the one-loop two-point renormalization, and in particular outline one-loop renormalization of the Newton's constant.

\end{abstract}
\newpage

%%%%%%%%%%%%%%%%%%%%%%%%%%%%%%%%%%%%%
%%%%%%%%%%%%%%%%%%%%%%%%%%%%%%%%%%%%%
\section{Introduction}
%%%%%%%%%%%%%%%%%%%%%%%%%%%%%%%%%%%%%
%%%%%%%%%%%%%%%%%%%%%%%%%%%%%%%%%%%%%

 There have been several different efforts over an extended period to quantize gravity \cite{Stelle:1976gc,Antoniadis:1986tu,Weinberg3,Reuter:1996cp,Odintsov:1990qq,Barvinsky:1993zg,VanNieuwenhuizen:1981ae,Bern:2011qn,Ashtekar:1986yd,Thiemann:2007zz,Ambjorn:2012jv,Calcagni:2012hb,Donoghue:2015hwa} and
the search for the true degrees of freedom of a gravitational theory has a similarly long history. Notably, it was stated in \cite{York:1972sj} that the true dynamical degrees of freedom of the gravitational field are associated with certain hypersurfaces of the spacetime under consideration. Several related discussions can be found in \cite{Moncrief:1989dx,Fischer:1996qg,Gay-Balmaz:2014ena}. 

The holographic characteristic is also present in the 't Hooft's observation on the black hole's degrees of freedom and AdS/CFT type dualities. With the motivation of better understanding the AdS/CFT type dualities and inspiration from the work of \cite{Sato:2002kv}, a variation of dimensional reduction technique - ``dimensional reduction to a hypersurface of foliation" - was introduced in \cite{Park:2013vpa}. In general, one gets to study lower dimensional subsectors of the original theory by manually reducing the theory to lower dimensions. (This remains true for dimensional reduction to a hypersurface of foliation in general except that the reduction is not manual but induced by gauge-fixing.)
The ADM formalism utilized in the manner in \cite{Park:2013vpa} led to complete gauge fixing of (almost) all undynamical fields, which in turn led to {\em ``spontaneous"} (as opposed to manual) dimensional reduction of the original theory to a hypersurface \cite{Park:2014tia}. Unlike the manual dimensional reduction, one does not lose the physical degrees of freedom of the original theory since one still covers the whole physical sector of the original theory, which in turn is due to the fact that the spontaneous reduction is brought along by gauge-fixing.  

It was observed in \cite{Park:2014tia} that the physical state space reduces to 3D at the level of the classical analysis.\footnote{
This holographic property may appear different from that of AdS/CFT type dualities in that the gauge theory degrees of freedom in the latter may seem alien to the original gravity system whereas the ``dual" degrees of freedom in the former are ``akin" to the original gravity theory. This is not the case: in the both the cases the dual holographic degrees of freedom are part of the original gravity theory.	The real difference is that in the present case they are directly visible in the sense that they do not require any transformation to become visible whereas the gauge degrees of freedom in AdS/CFT become visible after a certain ``dualization process" involving the Hamilton-Jacobi formalism \cite{Sato:2002kv}\cite{Hatefi:2012bp}.}  (A related linear-level analysis  can be found in an earlier work of \cite{Higuchi:1991tk}.) A canonical operator quantization was then outlined.
 Based on the mathematical observation discussed in detail in \cite{Park:2014qoa} and \cite{Park:2015qxa}, we believe that the reduction of the physical states is a property of a globally hyperbolic spacetime and is independent of a gauge choice. The present work will also present a slightly improved understanding of the issue of gauge-choice independence of the reduction; we will have more in the conclusion.

Comparison of the present quantization scheme with the pre-existing approaches should be useful. Various reductions in the true degrees of freedom were reported in the past in \cite{York:1972sj,Moncrief:1989dx,Fischer:1996qg,Gay-Balmaz:2014ena}, all of which employed the usual 3+1 splitting in which the genuine time coordinate was separated out. For instance, it was observed in \cite{York:1972sj} that the spacelike hypersurface specified up to a conformal factor can be taken as the true degrees of freedom. 
	In \cite{Moncrief:1989dx,Fischer:1996qg} it was shown that the reduced Hamiltonian was given by the volume of a certain hypersurface after Hamiltonian reduction was carried out on a class of 4D manifolds with certain topological restrictions. The present and related works \cite{Park:2014tia,Park:2014qoa,Park:2015qxa,Park:2014noa} tackle not only the issue of the true degrees of freedom but also the issue of quantization of gravity. Also,  a 3+1 splitting with one of the spatial directions separated out was employed. 
	For quantization, one should deal with the constraints, the so-called spatial diffeomorphism and Hamiltonian constraints. How to solve the diffeomorphism constraint has been one of the major obstacles in the gravity quantization.   
	 In our works in which the Lagrangian formalism was adopted, these constraints were called the shift vector and lapse function constraints, respectively, and were explicitly solved. The implication of the solution of the shift vector constraint has been brought out in \cite{Park:2014qoa,Park:2015qxa} by foliation theory. Throughout these works the strategy for reduction has been removal of all of the unphysical degrees of freedom from the external states. One of the key observations for the reduction was the fact that the residual 3D gauge symmetry can be employed to gauge away the non-dynamical fields such as the lapse function and shift vector. This is in addition to the standard bulk gauge-fixing. A detailed discussion of the 3D residual symmetry can be found in \cite{Park:2014noa}.

Coming back to the present scheme of qunatization, there exist different quantization approaches depending on the level of covariance. One may maintain only the 3D covariance throughout by adopting, as in \cite{Park:2014tia}, the canonical operator method and the corresponding path integral. In this work, we adopt a method that is ``largely" covariant:
the off-shell degrees of freedom will be four-dimensional whereas the on-shell physical states will explicitly be taken as three-dimensional.
More specifically, we first compute the counterterms for the Green's functions by the standard but refined background field method and obtain the 4D covariant 1PI effective action. We employ a traceless propagator to properly take care of the unphysical mode, the trace piece of the fluctuation metric.\footnote{As explicitly and extensively analyzed in a more recent work \cite{Park:2015xoa}, the removal of the trace piece can be achieved through {\em gauge-fixing}.} (Our method is different from that of \cite{Mazur:1989by}.) This turns out to be a crucial point missed in the literature: in the literature the covariance of the terms in the 1PI action was presumed and their coefficients were subsequently determined. Our analysis reveals that the presence of the trace piece is incompatible with the covariance. 
We consider amplitudes in which the external states are physical states, thus ``three-dimensional." The effective action corresponding to such amplitudes can be obtained by setting the fields in the original effective action to be the ``three-dimensional" physical fields.

When computing the off-shell 1PI action, we apply the background field method (BFM) in a manner that manifests the 4D covariance in spite of the fact that the action has been expanded around a fixed physical background (such as a flat spacetime). The conventional way of applying the BFM leads to non-covariant forms of the counterterm and this problem is known in the literature. (See, e.g., ch. 3 of \cite{Buchbinder}.) The refined application of the BFM is one of the new ingredients of the present work and the precise method will be explained below. Although the BFM is applied in such a way to restore the 4D covariance, the resulting counterterms still turn out to be {\em inexpressible} in terms of the covariant quantities. Interestingly, this pathology seems to originate from the problem noticed in \cite{Gibbons:1978ac}\cite{Mazur:1989by} some time ago: the trace piece of the fluctuation metric. (See further comments below.) After computing the off-shell 1PI action, we distinguish the renormalization of the S-matrix and that of the general off-shell Green's functions, and narrow down to the former. Choosing only the physical states to be the external legs corresponds to setting the fields in the effective action to be three-dimensional, which is according to the standard LSZ procedure. All of the Riemann tensor terms appearing in the effective action reduce to the well-known expression in terms of the Ricci tensor and metric. In other words, the whole effective action can now be expressed in terms of the 3D Ricci tensor and metric. With this one can introduce, as well known \cite{'tHooft:1973us}, a metric field redefinition to absorb the counterterms and thereby establish renormalizability of the original 4D action. (Therefore, the renormalizability is more complicated than the usual cases in that it requires a field redefinition that is unnecessary in a simply renormalizable theory in which only shifts in the parameters are required.)

Although the refined application of the BFM, which led to a (supposedly) clear understanding of the problem encountered in \cite{Park:2014noa} with regards to why the analysis therein was not compatible with the 4D covariance, was initially expected to yield a covariant effective action, it turns out it does {\em not}. The root of the problem is the pre-loop divergence, observed in \cite{Kuchar:1970mu}\cite{Gibbons:1978ac}\cite{Mazur:1989by}, caused by the presence of an unphysical gauge-mode, the trace piece of the fluctuation metric. While the issue must be genuine, it has not been clear in the literature where in the actual computations the pathology arises. In other words, it appears that one bypasses the problem in the usual manner of applying the BFM. In the refined BFM, in which expansion of the action around a fixed background is considered, one actually sees the problem arise in a manner not observed before as we will analyze in detail below. The pathology can be cured and the crux of the solution lies in removal of the unphysical degrees of freedom: once the trace part of the fluctuation metric is separated out and removed,\footnote{{The necessity of imposition of the traceless condition with the de Donder gauge was previously mentioned in \cite{Ortin} as we have recently become aware of.}} the 4D covariance results.

\vspace{.3in}

The rest of the paper is organized as follows.
\vspace{.1in}

We start by reviewing two things: determination of the physical states and the background field method. In section 2.1, the reduction of the physical states is shown without employing the full set of nonlinear de Donder gauge and therefore the present account is a slightly improved version of the reduction analyses of \cite{Park:2014tia} or \cite{Park:2015qxa}. The reduction of the physical states obviously raises a 4D covariance issue which is one of the two covariance issues that we address in this work. As for the background field method, we pay careful attention to the way that the covariant background field method reorganizes the Feynman diagrams associated with the action expanded around a given background. The key feature of our refined background field method is the ``double"-shift given in \rf{ds}. As a matter of fact, essentially the same technique had been employed in \cite{Antoniadis:1995fc}. We illustrate all these and subsequent ideas by revisiting one-loop two-point amplitudes \cite{'tHooft:1974bx,Deser:1974cz,Goroff:1985th}.  
For renormalization, we first follow the off-shell 4D covariant method and compute the off-shell effective action in section 3; we employ the traceless propagator in order to address the other covariance issue and thereby effectively remove the trace piece of the fluctuation metric. As demonstrated in section 3.1, one can easily see the pathology of the trace piece by examining the ghost sector covariance. (See also Appendix B.) One gets a variety of 4D-covariant counterterms prior to using the set of constraints laid aside until that point. Once the gauge-fixing and the lapse and shift constraints are used,\footnote{The gauge-fixing method of the present work has recently been generalized to a black hole background in \cite{Park:2015xoa}. } the effective action reduces to 3D, as we will show in section 4. All of the infinite number of couplings will become inessential since it is possible to absorb them by a field redefinition. 
We then outline renormalization of the coupling constant. The final section has our conclusion and future directions.

%%%%%%%%%%%%%%%%%%%%%%%%%%%%%%%%%%%%%
%%%%%%%%%%%%%%%%%%%%%%%%%%%%%%%%%%%%%
\section{Setup for holographic quantization}
%%%%%%%%%%%%%%%%%%%%%%%%%%%%%%%%%%%%%
%%%%%%%%%%%%%%%%%%%%%%%%%%%%%%%%%%%%%

In this section we set the stage for 4D-covariant holographic quantization, a task that will be taken up in the next section. Here we first review the analysis carried out in the previous works \cite{Park:2014tia,Park:2014noa,Park:2015qxa} in which the physical states were identified. Afterwards we review the BFM with an emphasis on how to utilize it in the setup of the action expanded around a given background in such a way as to not lose the 4D covariance. 
We stress the necessity of employing the traceless propagator \cite{Ortin}\cite{Park:2014tia} and double-shift technique, \rf{ds} \cite{Antoniadis:1995fc}\cite{Park:2014noa}. We also solve the puzzle\footnote{The one-loop two-point counterterms were computed in \cite{Park:2014noa} with the following field redefinition introduced by \cite{Capper:1973pv}:
\bea
q^{\a\b}\equiv \sqrt{-g}g^{\a\b};  \la{qab}
\eea 
below we repeat the amplitude/counterterm computation of \cite{Park:2014noa} without this field redefinition. In one of the footnotes of \cite{Park:2014noa}, we speculated that it might be this field redefinition that causes the problem. This is not the case: we have repeated the analysis without \rf{qab} and encountered the same puzzle. The upshot of the solution of the puzzle is that one must apply the BFM in the manner explained below and remove the trace part of the fluctuation metric.} encountered in \cite{Park:2014noa} in which the one-loop counterterms for the action expanded around a flat background were not expressible in terms of covariant quantities.

It may be useful to describe the strategy in a little more detail at this point to remain oriented. The BFM that we employ is the standard one with a refined interpretation. In the standard BFM, one considers
\bea
g_{\m\n}\equiv  h_{\m\n}+{g}_{{}_B\m\n}\quad 
\eea 
and integrates out the fluctuation field $ h_{\m\n}$ to get a covariant 
expression of ${g}_{{}_B\m\n}$. The 4D covariance is maintained, which is a great advantage for many things such as establishing renormalizability.  

%Let us suppose for now that renormalizability of gravity is not an issue. 
However, if one considers perturbation around a fixed physical background $g_{0\m\n}$ (such as, e.g., a flat or black hole spacetime), one expands the theory around $g_{0\m\n}$:  
\bea
g_{\m\n}\equiv  h_{\m\n}+g_{0\m\n}
\eea
One can compute various loop diagrams and will encounter divergences. Then one can determine the counterterms by making another shift:
\bea
h_{\m\n}\ra h_{\m\n}+g_{{}_B\m\n}
\eea
and integrating out $h_{\m\n}$ with the fields $g_{{}_B\m\n}$ serving as external lines.
Indeed, this was how the one-loop counterterms for 2-point amplitudes were obtained in \cite{Park:2014noa} with $g_{0\m\n}$ a flat background. 
The counterterms were not covariant and could not be expressed in terms of covariant quantities. 

To solve the puzzle (see also the comments below \rf{pothis}), we apply the BFM by considering the following shift \cite{Antoniadis:1995fc}\cite{Park:2014noa}
\bea
g_{\m\n}\equiv  h_{\m\n}+(g_{{}_B\m\n}+g_{0\m\n})   \la{ds}
\eea
and basically treating $g_{{}_B\m\n}+g_{0\m\n}$ as one piece in the manner explained in detail below. The reason for considering this form of the shift is two-fold. We need to establish renormalizability, and for that the background covariance is absolutely essential. Secondly, renormalization of the gravity will involve perturbative analysis around a given background for which again our refined BFM will be useful. In other words, we apply the BFM in the way we have applied here because we want to accomplish two things at the same time: establishing renormalizability and an actual perturbative analysis around a given background.

%%%%%%%%%%%%%%%%%%%%%%%%%%%%%%%%%%%%%
\subsection{determination of physical states \la{phystatereview}}
%%%%%%%%%%%%%%%%%%%%%%%%%%%%%%%%%%%%%

With any theory, determination of the physical states is an essential part of the quantization procedure. It is especially true for the present case because we propose to focus on the Feynman diagrams with the {\em physical} external legs. We review the analysis in \cite{Park:2014tia} by following a slightly different route; the result will be critical in the next section.

Consider the 4D Einstein-Hilbert action 
%%%
\bea
S=\int d^4 x \sqrt{-g}\;R  \la{unsplit}
\eea
%%%
and quantization in the operator formalism.
We split the coordinates into
%%%
\bea
x^\m\equiv (y^m,x^3) \la{coord}
\eea
%%%
where $\m=0,..,3$ and $m=0,1,2$. 
By parameterizing the 4D metric \cite{Arnowitt:1962hi}\cite{Poisson} in the the 3+1 split form, one gets 
%%%
\bea
g_{\m\n}=\left(
\begin{array}{cc}
\g_{mn} & N_{ m} \\
&\\
N_{ n} &  n^2+\g^{mn}N_{m} N_{ n} 
\end{array}
\right)\quad,\quad
g^{\m\n}=\left(
\begin{array}{cc}
\g^{mn}+\fr1{n^2}N^m N^n & -\fr1{n^2}N^{ m} \\
&\\
-\fr1{n^2}N^{ n} &  \fr1{n^2} 
\end{array}
\right)
\eea
%%%
where $n$ and $N_{m}$ denote the lapse function and shift vector respectively.
The action takes
%%%
\bea
S=\int d^4 x\;n\sqrt{-\g} \left(R^{(3)}+K^2-K_{mn}K^{mn}\right)
\la{1p3act}
\eea
%%%
with the second fundamental form given by
%%%
\be
K_{mn}=\fr1{2n}\left(\mathscr{L}_{\pa_{3}} \g_{mn}-{\nabla}_m N_{n}
         -{\nabla}_n N_{ m} \right),\qquad K=\g^{mn}K_{mn}.
\la{K4defqq}
\ee
where $\mathscr{L}_{\pa_{3}}$ denotes the Lie derivative along the vector field $\pa_{x^3}$ and $\N_m$ is the 3D covariant derivative constructed out of $\g_{mn}$. The shift vector and lapse function are non-dynamical and their field equations are 
%%%
\bea
{\N}_m (K^{mn}-\g^{mn} K)=0  \la{Ncon}
\eea
%%%
\bea
R^{(3)}-K^2+K_{mn}K^{mn}=0  \la{ncon}
\eea
%%%
The trace part of the metric field equation reads $R^\4=R^{(3)}+K^2-K_{mn}K^{mn}=0$. Combining this with \rf{ncon}, one has
\bea
R^\3=0=K^2-K_{mn}K^{mn}
\eea
The second equality then leads to $K_{mn}K^{mn}=0$ which in turn implies 
\bea
K_{mn}=0
\eea 
after Wick rotation and selection of a gauge $K=0,N_m=0$. To recapitulate, since all of the gauge freedom is used, the number of the physical components is fixed, prior to considering the lapse constraint, to be two. Any additional constraint that would reduce independent components, therefore, should not arise from the lapse constraint. The only way of ensuring this is the reduction in the coordinate dependence, which will make the $K_{mn}K^{mn}$ term "identically" vanish. At this point one can legitimately go to the Euclidean space by a Wick rotation; it follows \cite{Park:2014tia,Park:2015xoa} that $K_{mn}=0$.
Therefore, the physical states of a global spacetime that admits such a gauge choice are reduced to 3D.\footnote{Although $K=0$ is motivated by one of the four nonlinear components of the de Donder gauge, it may be possible to view it as a gauge choice independent of the de Donder gauge. It is at the quantum level of the fluctuation fields that one considers Wick rotation; $K_{mn}K^{mn}=0$ will admit 4D configurations at the classical level. (There is a subtlety: the Schwarzschild solution, for example, does not satisfy $K_{mn}K^{mn}=0$. This issue has been carefully addressed in \cite{Park:2015xoa}.) More refined and generalized analysis of the present subsection can be found in \cite{Park:2016zgt}.} A similar result was obtained in \cite{Higuchi:1991tk} at the linear level some time ago.

%%%%%%%%%%%%%%%%%%%%%%%%%%%%%%%%%%%%%
\subsection{computation of counterterms through BFM}
%%%%%%%%%%%%%%%%%%%%%%%%%%%%%%%%%%%%%

Although the BFM is usually employed for maintaining covariance at each step of the renormalization procedure, it does not need to be limited to that purpose. The method can be employed regardless of the covariance just to efficiently compute the effective action (especially its counter term portion), and this was how the method was employed in \cite{Park:2014noa}. However, covariance is an invaluable asset in establishing renormalizability.     
Below we discuss a way of applying the BFM in a manner that preserves covariance even after expanding the action around a fixed background. The analysis leads to a solution of the puzzle (after taking one additional subtlety into account) encountered in \cite{Park:2014noa}. We realize in retrospect that 4D covariance was sought in \cite{Park:2014noa} in a formulation where the covariance was obscured. We stress that there is a way to apply the BFM to accomplish both goals of maintaining the covariance and efficiently computing the counterterms. 

The additional subtlety is the one addressed in \cite{Kuchar:1970mu}, \cite{Gibbons:1978ac} and \cite{Mazur:1989by}. Although the pathology that these authors observed has not been detected in the standard BFM, it actually does seem to arise in our loop computation.\footnote{ Unlike in non-gravitational cases, the standard application of BFM becomes problematic due to the fact that a perturbation theory around a zero metric is not well defined.} The pathology manifests in such a way that 
the presence of an unphysical mode makes the path integral
divergent and non-covariant. We will see that omission of the trace piece of the fluctuation metric through gauge-fixing brings the anticipated 4D covariance at the level of the off-shell 1PI effective action.

Let us illustrate the pertinent issues by taking the scalar $\l \z^3$ theory,
%%%
\bea
\cL=-\fr12\pa_\m \z \pa^\m \z-\fr{\l}{3!}\z^3  \la{ztheory}
\eea
%%%
In the standard background field method, one computes the counterterms by shifting 
%%%
\bea
\z\ra \z+\z_B  \la{zzb}
\eea
%%%
\begin{figure}[tbp]
\centering 
\includegraphics[width=1\textwidth,trim=50 490 0 75,clip]{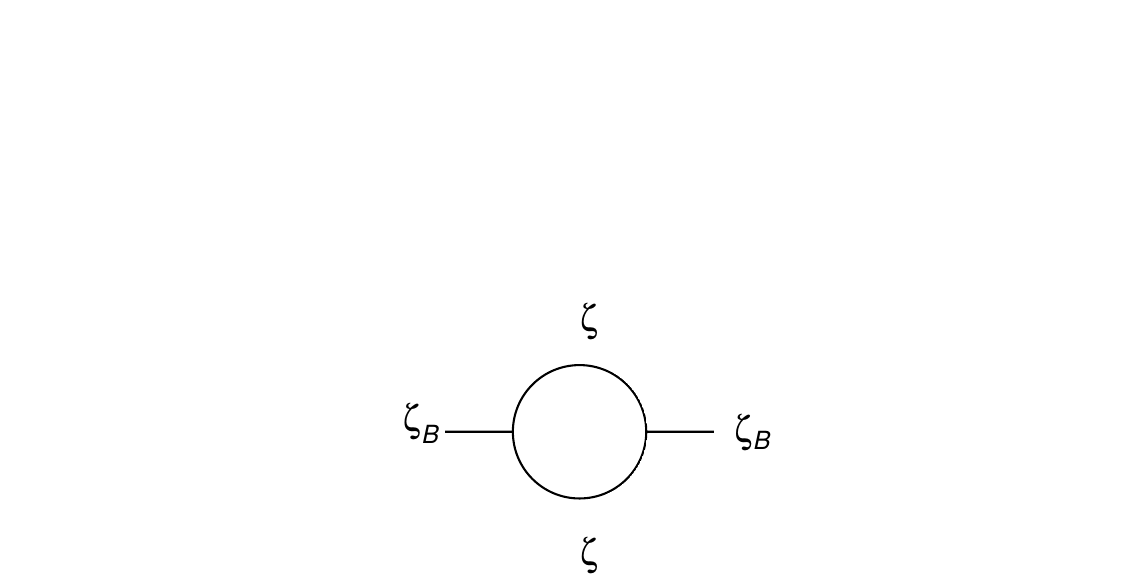}
\caption{$\l \z^3$ theory, BFM for zero vacuum}
\la{scalar2pt}
\end{figure}
%%%
\ni Once one integrates out $\z$ with the background field $\z_B$ placed in the external lines as shown, e.g., in Fig \ref{scalar2pt}., the counterterms for the one-loop two-point diagram are obtained.  
We emphasize that the counterterms thus obtained are suitable for cancelling the one-loop divergence produced by perturbation around the {\em zero-vacuum $\z=0$}.

One may consider a nonzero constant\footnote{We consider a constant vacuum for simplicity. The discussion does not depend on the constancy of the vacuum in any essential way.} vacuum $\z=\z_0$ and expand the theory \rf{ztheory} around this vacuum: 
%%%
\bea
\cL 
    &=& -\fr12\pa_\m \z \pa^\m \z-\fr{\l}{3!}(\z^3+3\z_0\z^2+3\z_0^2\z+\z_0^3)
\la{z0theory}
\eea
%%%
This Lagrangian will have one-loop divergence, and one can again compute the counterterms by employing the BFM; shift the action \rf{z0theory} around $\z=\z_B$
%%%
\bea
\cL=-\fr12\pa_\m (\z+\z_B+\z_0) \pa^\m (\z+\z_B+\z_0)-\fr{\l}{3!}(\z+\z_B+\z_0)^3  \la{z0theoryde}
\eea
%%%
and integrate out the $\z$ field with the background field $\z_B$ becoming external lines. The computation of \cite{Park:2014noa} (and some works in the past) was done analogously. Not all is well since this way the 4D covariance was obscured.

We now get to the way of applying the BFM both to effectively compute the counterterms and to maintain covariance even when considering a nonzero background. Let us introduce
\bea
{\xi}_B=\z_B+\z_0
\eea
for convenience; \rf{z0theoryde} can be written as
%%%
\bea
\cL=-\fr12\pa_\m (\z+{\xi}_B) \pa^\m (\z+{\xi}_B)-\fr{\l}{3!}(\z+{\xi}_B)^3  \la{z0theoryde2}
\eea
%%%
\begin{figure}[tbp]
\centering 
\includegraphics[width=1.3\textwidth,trim=70 650 -90 75,clip]{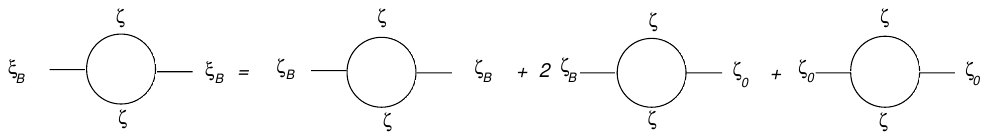}
\caption{$\l \z^3$ theory, BFM for nonzero vacuum}
\la{scalar4figs}
\end{figure}
Now suppose we use \rf{z0theoryde} but now integrate the fluctuation field $\z$ such that $\z_B+\z_0$ becomes the external fields. (This case can of course be viewed as the case of \rf{z0theoryde2} and \rf{zzb} with ${\xi}_B$ replaced by $\z_B+\z_0$.) As shown in Fig \ref{scalar4figs}., integrating out the scalar field with $\xi_B$ as external fields corresponds to the sum of three diagrams now with $\z_B,\z_0$ serving as external legs. The necessity of considering the second and third diagrams on the right-hand side of Fig \ref{scalar4figs} is obvious from the forms of the vertices appearing in \rf{z0theory}.

This simple observation on the external lines provides the solution to the puzzle encountered in \cite{Park:2014noa}.
Let us take the actual system of our interest, the gravity action. One can shift the metric
\bea
g_{\m\n}\equiv  h_{\m\n}+\tilde{g}_{{}_B\m\n}\quad \mbox{where}\quad \tilde{g}_{{}_B\m\n}\equiv \vf_{{}_B\m\n}+g_{0\m\n}
\la{pothis}
\eea
and integrate out $h_{\m\n}$ with the background field $\tilde{g}_{{}_B\m\n}$ appearing on the external lines (see Fig.\ref{graviBFM}). The resulting 1PI effective action will be covariant function of $\tilde{g}_{{}_B\m\n}(\equiv \vf_{{}_B\m\n}+g_{0\m\n})$. (Since this was not how the analysis was carried out in \cite{Park:2014noa}, the covariance got obscured.) 
The resulting action would be covariant in normal (i.e., non-gravitational) circumstances. However, there is another subtlety in a gravitational case: the presence of the unphysical mode, in particular the trace part of the fluctuation metric. Its presence interferes with the covariance as we will see shortly. It is possible to carry out the perturbation theory by adopting a traceless propagator, one of whose effects is removal of the trace part, which is essentially a gauge-fixing \cite{Park:2015xoa}, from the effective action. We turn to the details of this analysis.

%%%%%%%%%%%%%%%%%%%%%%%%%%%%%%%%%%%%%
%%%%%%%%%%%%%%%%%%%%%%%%%%%%%%%%%%%%%
\section{4D covariant off-shell effective action}
%%%%%%%%%%%%%%%%%%%%%%%%%%%%%%%%%%%%%
%%%%%%%%%%%%%%%%%%%%%%%%%%%%%%%%%%%%%

\begin{figure}[tbp]
\centering 
\includegraphics[width=1.1\textwidth,trim=50 640 -10 15,clip]{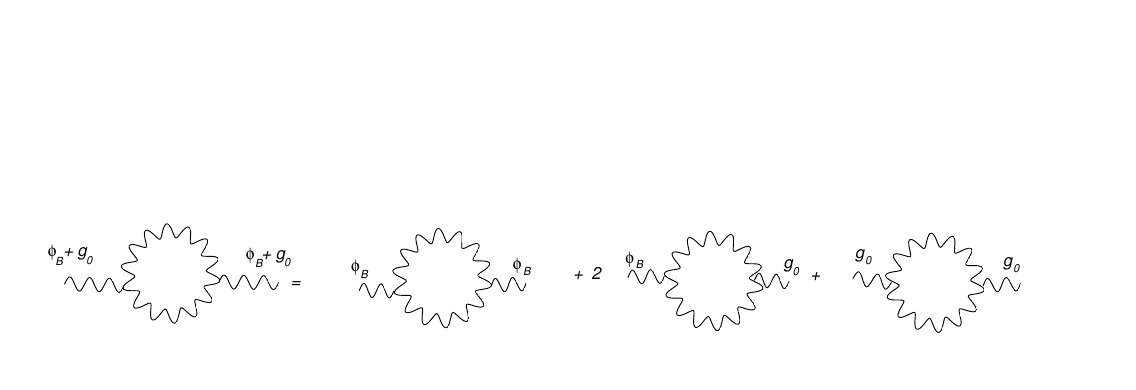}
\caption{BFM for graviton one-loop (spacetime indices suppressed)}
\la{graviBFM}
\end{figure}

With all of the ingredients reviewed, we can now carry out the one-loop renormalization.
One can set aside the fact that more and more counterterms arise as the order of the loop becomes higher, and just patiently obtain all of the counterterms at each order. Suppose one has obtained the 4D covariant 1PI action to the desired order. One can then substitute the physical state constraints. (For perturbation, we should linearize\footnote{This is in the spirit of the footnote $\dagger$ on page 33 (ch. 15) of \cite{Weinberg2}.} the constraint equations but still keep the 3D characteristics.) The resulting expression will contain all the physical pieces of information. Doing things off-shell at the path integral level would just make the form of the 4D 1PI action more complicated compared with \cite{Park:2014tia,Park:2014noa}, in which one carries out things entirely in the 3D sense. However, this additional complexity will be absorbed by a field redefinition that will be implemented after substituting the constraints in the effective action.

Consider the Einstein action 
%%%
\bea
S=\fr1{\k^2}\int d^4 x \sqrt{-g}\;R
  \la{unsplit}
\eea
%%%
where $\k^2= 16\pi G$ with $G$ being Newton's constant; $\k^2$ will be suppressed in this section.
By shifting the metric according to 
\bea
g_{\m\n}\equiv  h_{\m\n}+\tilde{g}_{{}_B\m\n}\quad \mbox{where}\quad \tilde{g}_{{}_B\m\n}\equiv \vf_{{}_B\m\n}+g_{0\m\n},\quad g_{0\m\n}=\eta_{\m\n}
\eea
one obtains \cite{Goroff:1985th}\footnote{{The convention of \cite{Goroff:1985th} is such that
$R_{\m\n}\equiv R^\k{}_{\m \n\k}$ whereas ours is $R_{\m\n}\equiv R^\k{}_{\m \k\n}$. The sign of $h^{\a}{}_{\a}h_{\b\g}\Rt^{\b\g}$-term here is opposite to that of \cite{Goroff:1985th} for this reason. (Also it appears that the signs of the last two terms in the corresponding equation in \cite{Goroff:1985th} have typos.)}}
\bea
 \cL = \sqrt{-\gt}\Big( -\fr12\tilde{\N}_\g h^{\a\b}\tilde{\N}^\g h_{\a\b}+\fr14 \tilde{\N}_\g h^{\a}_\a \tilde{\N}^\g h^{\b}_\b  +h_{\a\b}h_{\g\d}\Rt^{\a\g\b\d}\la{gravcubcov}
\eea
\[-h_{\a\b}h^{\b}{}_\g \Rt^{\k\a\g}{}_{\k}
{ -} h^{\a}{}_{\a}h_{\b\g}\Rt^{\b\g}-\fr12 h^{\a\b}h_{\a\b}\Rt
+\fr14  h^{\a}_\a  h^{\b}_\b \Rt +\cdots\Big)  
\]
%%%%%%%%%
The fields with a tilde are the background quantities;
\begin{figure}[tbp]
\centering 
\includegraphics[width=.9\textwidth,trim=0 620 0 65,clip]{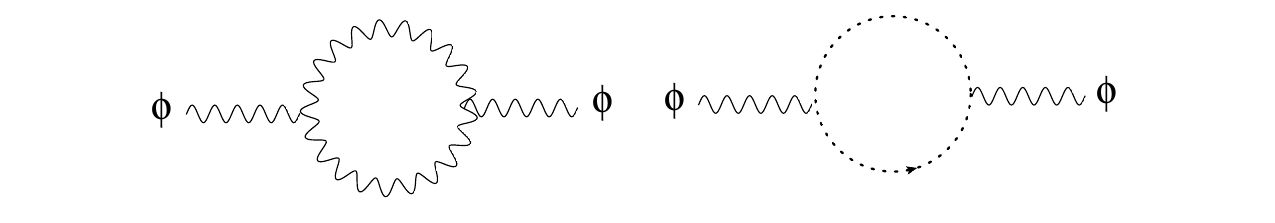}
\caption{(a) graviton loop  (b) ghost loop (spacetime indices suppressed)}
\la{gravgh}
\end{figure}
the ghost action is given by
\bea
{\cal L}_{\mbox{gh}}&=&  -\tilde{\N}^\n \Cb^\m \tilde{\N}_\n C_\m  {+\tilde{R}_{\m\n}}\bar{C}^\m C^\n +\cdots 
\eea
Here and \rf{gravcubcov} only the one-loop pertinent terms have been kept.

\subsection{removal of trace part of  metric}

As mentioned previously, several authors had pointed out in the past the problem caused by the presence of unphysical components of the fluctuation metric.
One can see the manifestation of the problem in the present setup by taking a detailed look at the kinetic part of the ghost sector:
\bea
{\cal L}_{\mbox{gh}}&=&  -\sqrt{-\gt}\;\tilde{\N}^\n \Cb^\m \tilde{\N}_\n C_\m
\la{ghkin}
\eea
Its contributions to the effective action, in particular the counterterms, are expected to come out in background-covariant forms. As we have pointed out, the usual analysis is formal and does not seem to properly address the pathological divergence observed in \cite{Gibbons:1978ac} and \cite{Mazur:1989by}.
We show below that the presence of the trace piece of the fluctuation metric indeed causes problems: it ruins the covariance.

For the complete evaluation of the counterterms from \rf{ghkin}, one should integrate out the ghost fields with the {\em full} propagator. Instead of attempting the full evaluation at one stroke, one may perturbatively produce the full result by carrying out path integrals around a flat spacetime to the desired order. In other words, if we perturbatively produce the result, say, to the second order in the fields, we are ``probing" the full result to the second order. (One can of course go to higher orders.) Let us first expand the metric $\gt_{\m\n}$ around a flat spacetime:
\bea
\gt_{\m\n}\equiv \vf_{\m\n}+\eta_{\m\n}
\eea
where we have omitted the letter $B$ from $\vf_{{}_B\m\n}$.
The kinetic ghost action \rf{ghkin} takes
\bea
{\cal L}_{\mbox{gh}}&=& -\Big[
\pa^\m \bar{C}^\n \pa_\m {C}_\n+\fr{1}{2}\vf \pa^\m \bar{C}^\n \pa_\m {C}_\n
-\tilde{\G}^\l_{\m\n}({ -C_\l} \pa^\m \bar{C}^\n+\bar{C}_\l\pa^\m {C}^\n  )\nn\\
&& -(\eta^{\n\b}\vf^{\m\a}+\eta^{\m\a}\vf^{\n\b})\pa_\b \bar{C}_\a \pa_\n {C}_\m
\Big]  \la{ghkinexp}
\eea
The counterterms for the diagrams in Fig.\ref{gravgh} (b) can be computed; naively, the result is anticipated to be 4D covariant. In other words, it will be given by the following expression expanded to the second order in fields:
\bea
a_1 \Rt^2+a_2 \Rt_{\m\n}\Rt^{\m\n} \la{covctrcomb}
\eea   
with some numerical coefficients $a_1,a_2$. It turns out that the result is {\em not} precisely of this form: the terms containing the trace (i.e., $\vf\equiv \eta^{\m\n}\vf_{\m\n}$) destroy the covariance. Once one sets $\vf=0$, which can be effectively achieved by employing the traceless propagator, the result takes the covariant form given in \rf{covctrcomb}.

The details are as follows. In the position space, the correlator to be computed is given by
\bea
-\fr12<\Big[\int \fr{1}{2}\vf \pa^\m \bar{C}^\n \pa_\m {C}_\n
-\tilde{\G}^\l_{\m\n}({ -C_\l}\pa^\m \bar{C}^\n+\bar{C}_\l\pa^\m {C}^\n  )
-(\eta^{\n\b}\vf^{\m\a}+\eta^{\m\a}\vf^{\n\b})\pa_\b \bar{C}_\a \pa_\n {C}_\m
\Big]^2>\nn\\
\eea
It can be grouped into
\bea
(I) &\equiv & -\fr12 \int\int \Big[ \fr{1}{2}\vf \eta^{\b\n}\eta^{\a\m} 
-\eta^{\n\b}\vf^{\m\a}-\eta^{\m\a}\vf^{\n\b})
\Big]\Big[ \fr{1}{2}\vf \eta^{\b'\n'}\eta^{\a'\m'} 
-\eta^{\n'\b'}\vf^{\m'\a'}-\eta^{\m'\a'}\vf^{\n'\b'})
\Big]\nn\\
&& \;\;\;\;\hspace{.5in}<(\pa_\b \bar{C}_\a \pa_\n {C}_\m) (\pa_{\b'} \bar{C}_{\a'} \pa_{\n'} {C}_{\m'})>  \nn\\
%%%%%%%%%%%%%%%%%%%%%%%%%%%%%%%%%%%%%%%
(II) &\equiv &{ -\fr12 (2)} \int\int \Big[ \fr{1}{2}\vf \eta^{\b\n}\eta^{\a\m} 
-\eta^{\n\b}\vf^{\m\a}-\eta^{\m\a}\vf^{\n\b})
\Big]\tilde{\G}^{\l'}_{\m'\n'}
 <(\pa_\b \bar{C}_\a \pa_\n {C}_\m)(-\pa^{\m'} \bar{C}^{\n'} {C_{\l'}}+\pa^{\m'} {C}^{\n'} \bar{C}_{\l'} )>  \nn\\
 (III) &\equiv & -\fr12 \int\int \tilde{\G}^\l_{\m\n}\tilde{\G}^{\l'}_{\m'\n'}
 <(-\pa^{\m} \bar{C}^{\n} {C_{\l}}+\pa^{\m} {C}^{\n} \bar{C}_{\l} )(-\pa^{\m} \bar{C}^{\n'} { C_{\l'}}+\pa^{\m'} {C}^{\n'} \bar{C}_{\l'} ) >
\eea
The counterterms can be isolated from the divergent parts; we found (as in \cite{Park:2014noa}, we employ a Mathematica package xAct`xTensor` for numerical tensor manipulations)
\bea
(I) &\Rightarrow&-\fr12 \fr{\G(\e)}{(4\pi)^2}\Big[
\fr2{15} (\pa_\a \pa_\b \vf^{\a\b})^2-\fr1{30} \pa^2 \vf (\pa_\a \pa_\b \vf^{\a\b})
-\fr1{15}\pa^2 \vf \pa^2 \vf   \nn\\
&&-\fr1{15}\pa^2 \vf^{\a\k}\pa_\k \pa_\s \vf_\a^\s+\fr{17}{60}\pa^2\vf_{\m\n}\pa^2 \vf^{\m\n}
\Big]\nn\\
(II)&\Rightarrow& -\fr12 \fr{\G(\e)}{(4\pi)^2}\Big[ -\fr12\pa^2 \vf_{\a\b}(x)\pa^2 \vf^{\a\b}  +\fr16 \pa^2\vf(x)\pa^2 \vf  -\fr16 \pa^{\a}\pa^{\b}\vf_{\a\b}\pa^2 \vf  \Big]\nn\\
(III)&\Rightarrow& -\fr12 \fr{\G(\e)}{(4\pi)^2}{ \Big[  
 \fr1{12}\pa^2 \vf_{\a\b}(x)\pa^2 \vf^{\a\b} -\fr16(\pa_\a \pa_\b \vf^{\a\b})^2+\fr13 \pa^2 \vf^{\a\k}\pa_\k \pa_\s \vf_\a^\s
\Big]} \nn\\
\eea
where the parameter $\e$ is related to the total spacetime dimension $d$ by
\bea
d=4-2\e
\eea
The total counter term Lagrangian from this sector is given by the sum of these contributions:
\bea
\D \cL &=&-\fr12 \fr{\G(\e)}{(4\pi)^2}\Big[ -\fr{2}{15}\pa^2\vf_{\m\n}\pa^2 \vf^{\m\n}+\fr{4}{15}\pa^2 \vf^{\a\k}\pa_\k \pa_\s \vf_\a^\s\nn\\
&&-\fr{1}{30}(\pa_\a \pa_\b \vf^{\a\b})^2
-\fr1{5} \pa^2 \vf \,\pa_\a \pa_\b \vf^{\a\b}
+\fr1{10}\pa^2 \vf \pa^2 \vf 
\Big] \la{ctrghpartI}
\eea
Upon setting $\vf=0$, this becomes
\bea
\D \cL =-\fr12 \fr{\G(\e)}{(4\pi)^2}\Big[ -\fr{2}{15}\pa^2\vf_{\m\n}\pa^2 \vf^{\m\n}+\fr{4}{15}\pa^2 \vf^{\a\k}\pa_\k \pa_\s \vf_\a^\s
-\fr{1}{30}(\pa_\a \pa_\b \vf^{\a\b})^2
\Big] \la{ctrghpartI2}
\eea
On account of
\bea
R^2 &\simeq& \pa_{\m}\pa_{\n}\vf^{\m\n}\,\pa_{\r}\pa_{\s}\vf^{\r\s}
\nn\\
R_{\a\b}R^{\a\b} &\simeq& \fr14\Big[\pa^2 \vf^{\m\n}\,\pa^2 \vf_{\m\n}-2\pa^2 \vf^{\a\k}\pa_\k \pa_\s \vf_\a^\s
+2(\pa_{\m}\pa_{\n}\vf^{\m\n})^2
 \Big]
\nn\\
\la{covctr}
\eea
\rf{ctrghpartI2} can be written 
\bea
\D \cL = -\fr1{2} \fr{\G(\e)}{(4\pi)^2}\Big[-\fr{8}{15}\Rt_{\a\b}\Rt^{\a\b}+\fr{7}{30}\Rt^2\Big]
\eea

\subsection{4D covariant 1-loop counterterms}

In the previous subsection we have examined a part of the ghost sector to illustrate how the pathology noted in 
\cite{Kuchar:1970mu,Gibbons:1978ac,Mazur:1989by} arises in actual computations. In this subsection we compute the graviton sector counterterms and complete the rest of the ghost sector, thereby obtaining the full counterterm action \rf{totalctr} below.

By expanding \rf{gravcubcov} one gets
\bea
\cL&=& -\fr12 {\pa}_\g h^{\a\b}{\pa}^\g h_{\a\b}+\fr14 {\pa}_\g h^{\a}_\a {\pa}^\g h^{\b}_\b  \nn\\
&& + \cL_{V_I}+\cL_{V_{II}}+\cL_{V_{III}} 
\eea
where
\bea
\cL_{V_I} &=&   \Big(2\eta^{\b\b'}\tilde{\G}^{\a' \g\a}- \eta^{\a\b}\tilde{\G}^{\a' \g\b'}\Big)\pa_\g h_{\a\b}\, h_{\a'\b'}  \nn\\
\cL_{V_{II}} &=& \Big[\fr12(\eta^{\a\a'}\eta^{\b\b'}\vf^{\g\g'}+\eta^{\b\b'}\eta^{\g\g'}\vf^{\a\a'}
+\eta^{\a\a'}\eta^{\g\g'}\vf^{\b\b'})\nn\\
&&-\fr14 \vf\, \eta^{\a\a'}\eta^{\b\b'}\eta^{\g\g'}-\fr12 \eta^{\g\g'}\eta^{\a'\b'}\vf^{\a\b}  \nn\\
&&+\fr14 (-\vf^{\g\g'}+\fr12 \vf \eta^{\g\g'})\eta^{\a\b}\eta^{\a'\b'}
\Big] \pa_\g h_{\a\b}\, \pa_{\g'}h_{\a'\b'}  \la{lv12}
\eea
These vertices come from the first line of \rf{gravcubcov}\footnote{The distingction between $\cL_{V_I} $ and $\cL_{V_{II}}$ is made for convenience in Mathematica coding.} whereas the vertex $\cL_{V_{III}}$ is just the second line:
\bea
\cL_{V_{III}} = \Big( h_{\a\b}h_{\g\d}\Rt^{\a\g\b\d}-h_{\a\b}h^{\b}{}_\g \Rt^{\k\a\g}{}_{\k}
{ -}h^{\a}{}_{\a}h_{\b\g}\Rt^{\b\g}-\fr12 h^{\a\b}h_{\a\b}\Rt
+\fr14  h^{\a}_\a  h^{\b}_\b \Rt +\cdots\Big)  \la{gver}  \nn\\
\eea
For the complete one-loop two-point amplitude one should consider the following correlator:
\bea
&&\fr{i^2}{2}< (\cL_{V_I}+\cL_{V_{II}}+\cL_{V_{III}})^2 >\nn\\
&=&\fr{i^2}{2}\Big(< (\cL_{V_I}+\cL_{V_{II}})^2 >+2< (\cL_{V_I}+\cL_{V_{II}})\cL_{V_{III}} >
+< \cL_{V_{III}}^2 > \Big) \la{gcor} \nn\\
\eea 
One can show that the obviously covaraint correlator $< \cL_{V_{III}}^2 >$ is given by
\bea
&&\hspace{-.7in}-\fr12<\Big(h_{\a\b}h_{\g\d}\Rt^{\a\g\b\d}-h_{\a\b}h^{\b}{}_\g \Rt^{\k\a\g}{}_{\k}
{ -}h^{\a}{}_{\a}h_{\b\g}\Rt^{\b\g}-\fr12 h^{\a\b}h_{\a\b}\Rt
+\fr14  h^{\a}_\a  h^{\b}_\b \Rt \Big)^2> \nn\\
\hspace{.2in}&=&-\fr12<\Big(h_{\a\b}h_{\g\d}\Rt^{\a\g\b\d}-h_{\a\b}h^{\b}{}_\g \Rt^{\k\a\g}{}_{\k}
-\fr12 h^{\a\b}h_{\a\b}\Rt
\Big)^2> 
\eea
where the first equality is due to our use of the traceless propagator.
After some algebra, one gets
\bea
-\fr12\fr{\G(\e)}{(4\pi)^2}\Big[-3\Rt_{\m\n}\Rt^{\m\n}-\Rt^2\Big]  \la{tgravi}
\eea
For the other two correlators we present here the computation employing the traceless propagator; the computation employing the traceful propagator can be found in one of the appendices.\footnote{In contrast use of the traceful propagator leads to non-covariant results as presented in one of the appendices.} The first correlator in the right hand side of \rf{gcor} is 
\bea
-\fr12 < (\cL_{V_I}+\cL_{V_{II}})^2 > 
&=& -\fr12 \fr{\G(\ve)}{(4\pi)^2}\Big[-\fr{23}{40}R^2+\fr{37}{20} R_{\a\b}R^{\a\b}\Big]
\eea
The result of the second correlator is
\bea
-\fr12 2< (\cL_{V_I}+\cL_{V_{II}})\cL_{V_{III}} >
&=& -\fr12\fr{\G(\ve)}{(4\pi)^2} R^2
\eea
The total graviton contribution therefore is
\bea
&&-\fr{1}{2}< (\cL_{V_I}+\cL_{V_{II}}+\cL_{V_{III}})^2 >\nn\\
&=& -\fr12\fr{\G(\e)}{(4\pi)^2}\Big[(-3+\fr{37}{20})\Rt_{\m\n}\Rt^{\m\n}+(-1-\fr{23}{40}+{1})\Rt^2\Big] \nn\\
&=& -\fr12\fr{\G(\e)}{(4\pi)^2}\Big[-\fr{23}{20}\Rt_{\m\n}\Rt^{\m\n}-{ \fr{23}{40}}\Rt^2\Big] \la{gravitot}
\eea

The analysis for the ghost sector is parallel. The total ghost contribution would be obtained by integrating out the ghost fields with the full propagator. It should be possible to perturbatively produce the result by carrying out path integrals around a flat spacetime to the desired order. To the quadratic order for example, we can compute the following correlator with the flat spacetime propagator:  
\bea
&&\hspace{-.4in} -\fr12<\Big\{\Big[
\fr{1}{2}\vf \pa^\m \bar{C}^\n \pa_\m {C}_\n
-\tilde{\G}^\l_{\m\n}(\pa^\m \bar{C}^\n { C_\l} -\pa^\m {C}^\n  \bar{C}_\l  )
-(\eta^{\n\b}\vf^{\m\a}+\eta^{\m\a}\vf^{\n\b})\pa_\b \bar{C}_\a \pa_\n {C}_\m
\Big] { -}R_{\m\n}\bar{C}^\m C^\n\Big\}^2>
  \la{totgh1}
\nn\\
\eea
Part of this computation has been carried out in the previous section. After some tedious steps, one gets, 
{
\bea	
&&2<\Big[\fr{1}{2}\vf \pa^\m \bar{C}^\n \pa_\m {C}_\n-\tilde{\G}^\l_{\m\n}(\pa^\m \bar{C}^\n { C_\l} -\pa^\m {C}^\n  \bar{C}_\l  )
-(\eta^{\n\b}\vf^{\m\a}+\eta^{\m\a}\vf^{\n\b})\pa_\b \bar{C}_\a \pa_\n {C}_\m \Big] R_{\m\n}\bar{C}^\m C^\n> \nn\\
&=& -\fr12 \fr{\G(\e)}{(4\pi)^2}\Big[
		{ }\fr13 \Rt^2\Big]
\eea
	}
	and
\bea
&&\hspace{-.4in} -\fr12<\Big[R_{\m\n}\bar{C}^\m C^\n\Big]^2>=-\fr12 \fr{\G(\e)}{(4\pi)^2}R_{\m\n}R^{\m\n}
\eea
One gets, after setting $\vf=0$,
{
\bea
\rf{totgh1}
%%%%%%%%%%%%%%%%%%%%%%%%%%%%%%%%%%%%%%%%%%%%%%%%
&=& -\fr12 \fr{\G(\e)}{(4\pi)^2}\Big[-\fr{8}{15}\Rt_{\a\b}\Rt^{\a\b}+\fr{7}{30}\Rt^2
{+}\fr13\Rt^2
+\Rt_{\m\n}\Rt^{\m\n}\Big]\nn\\
%%%%%%%%%%%%%%%%%%%%%%%%%%%%%%%%%%%%%%%%%%%%%%%%
&=& -\fr12 \fr{\G(\e)}{(4\pi)^2}\Big[ \fr{7}{15}\Rt_{\m\n}{ \Rt^{\m\n}} {+\fr{17}{30}}\Rt^2
\Big] \la{tgh}
\eea}
The total one-loop counterterms therefore are given by the sum of \rf{gravitot} and \rf{tgh}: 
\bea
\D \cL_{1loop} &=& -\fr12\fr{\G(\e)}{(4\pi)^2}\Big[(-\fr{23}{20}+\fr{7}{15})\Rt_{\m\n}\Rt^{\m\n}+({ -\fr{23}{40}}{ +\fr{17}{30}})\Rt^2\Big]
\Big] \nn\\
&=& -\fr12\fr{\G(\e)}{(4\pi)^2}\Big[-\fr{41}{60}\Rt_{\m\n}\Rt^{\m\n}{ -{ \fr{1}{120}}}\Rt^2\Big]\la{totalctr}
\eea
This result is different from the well-known result in \cite{'tHooft:1974bx} for the reasons already stated: the present result has been obtained by using the traceless propagator which in turn has been motivated to remedy the pathology observed \cite{Kuchar:1970mu,Gibbons:1978ac,Mazur:1989by} by gauging away the trace piece of the fluctuation metric.

%%%%%%%%%%%%%%%%%%%%%%%%%%%%%%%%%%%%%
%%%%%%%%%%%%%%%%%%%%%%%%%%%%%%%%%%%%%
\section{Renormalization of S-matrix}
%%%%%%%%%%%%%%%%%%%%%%%%%%%%%%%%%%%%%
%%%%%%%%%%%%%%%%%%%%%%%%%%%%%%%%%%%%%

We have established in the previous section that the counterterms for the one-loop two-point amplitude can be expressed in terms of 4D covariant quantities. The renormalization procedure can be carried out to any order once only the physical external states are considered, since then the Riemann tensors in the effective action reduce to 3D and can be expressed in terms of the 3D Ricci tensor and metric. With the status of the renormalizability settled, it is only a matter of explicitly conducting the procedure. Here, we revisit one-loop renormalization of the metric that is accompanied by a field redefinition and outline one-loop renormalization of the coupling constant.\footnote{At one-loop, the Riemann appears in the form of $R_{\m\n\r\s}R^{\m\n\r\s}$ and there is a well-known identity:
\bea
R_{\m\n\r\s}R^{\m\n\r\s}-4R_{\m\n}R^{\m\n}+R^2=\mbox{total derivative} \la{Riemannsqid}
\eea
The renormalizability can be established without making use of the reduction of the 1PI action to 3D. However, this is special for one-loop; for higher loops one will need to make use of it. 
}

\subsection{reduction of 1PI action}

In this subsection, we discuss one of the central steps for establishing higher loop renormalizability, i.e., reduction of the 4D covariant effective action to 3D. The 1PI action in the previous section has been obtained by considering off-shell external states. Whereas it is not necessary to distinguish renormalization of the off-shell effective action from that of the S-matrix in a non-gravitational gauge theory (since it is renormalizable at the off-shell level anyway), this is not the case for a gravity theory. We show that Einstein gravity is renormalizable in a covariant approach (despite of the well-known fact that there appear infinite number of counterterms) once one considers only the physical external states.

Consider physical external states; as we have reviewed in section \ref{phystatereview}, they are three-dimensional. We can impose various constraints at the level of the 1PI effective action. (For this, we find the general field theory accounts in ch. 7 and 10 of \cite{Zinn-Justin} useful.) Although it should be possible to keep the discussion more abstract at the level of general LSZ reduction procedure by following, e.g., \cite{Sterman} we will instead consider a specific example.

We illustrate the idea by analyzing one of the terms in the ghost sector action \rf{ghkinexp}:
\bea
-\eta^{\n\b}\vf^{\m\a}\pa_\b \bar{C}_\a \pa_\n {C}_\m
\eea
The divergence of the one-loop two-point correlator from this vertex is given by
\bea
&&\int \fr{d^4x_1}{(2\pi)^4}\fr{d^4x_2}{(2\pi)^4}e^{-il_1\cdot x_1}e^{-il_2\cdot x_2}<h_{\m\n}(x_1)h_{\r\s}(x_2)>_{\mbox{{1-loop div}}} \nn\\
&=& -\fr12\int \fr{d^4x_1}{(2\pi)^4}\fr{d^4x_2}{(2\pi)^4}e^{-il_1\cdot x_1}e^{-il_2\cdot x_2}<h_{\m\n}h_{\r\s}\Big[\eta^{\n\b}h^{\m\a}\pa_\b \bar{C}_\a \pa_\n {C}_\m\Big]^2>\nn\\
&=&\fr14 \fr{\d(l_1+l_2)}{(2\pi)^4}P_{\m\n}{}^{\k_1\k_2}P_{\r\s\k_1\k_2}\fr1{l_1^2l_2^2}(l_1^4)
\eea
where $l_1,l_2$ denote the off-shell 4-momenta.
The contribution to the S-matrix of the physical two-point is obtained by removing the external propagators through the standard LSZ reduction procedure:
\bea
\Rightarrow \fr14 [2\pi\d(0)]\fr{(2\pi)^3\d(\vec{p}_1+\vec{p}_2)}{(2\pi)^4}P_{mnrs}\;\vec{p_1}^2 
\la{smdiv}
\eea 
where $(m,n,r,s)$ denote the 3D polarization indices and $\vec{p_1}$ the on-shell 3-momentum.
Applying the BFM, the counterterms can be shown to be
\bea
\D \cL &=& -\fr12<\Big[\eta^{\n\b}\vf^{\m\a}\pa_\b \bar{C}_\a \pa_\n {C}_\m \Big]^2>_{1loop}\nn\\
&=&-\fr18\int d^4x \;\pa^2\vf^{\m\n}(x^\m)\pa^2\vf_{\m\n}(x^\m)
\eea
Consider reduction of the coordinate dependence to 3D
\bea
\D \cL_{3D}=-\fr18\int dx^0\int d^3x \;\pa^2\vf^{mn}(x^i)\pa^2\vf_{mn}(x^i)
\eea
where $\pa^2$ is now three-dimensional.
One can easily check that the contribution of this counter term action to the S-matrix cancels the divergence given in \rf{smdiv} after Wick rotation. (Note that $\int dx^0=(2\pi)\d(0)$ \cite{Weinberg2}.)

%%%%%%%%%%%%%%%%%%%%%%%%%%%%%%%%%%%%%
\subsection{renormalization by field redefinition}
%%%%%%%%%%%%%%%%%%%%%%%%%%%%%%%%%%%%%

Once the effective action reduces to 3D, one can renormalize the metric by a field redefinition \cite{'tHooft:1973us}. For one-loop, it is the identity given in \rf{Riemannsqid} that brings the renormalizability through a field redefinition that we review now. One should follow similar steps in the 3D language in order to establish higher loop renormalizability.\footnote{In other words, one reduces things according to (``${}^\3$" denotes the 3D quantities)
\bea
R &\ra& R^\3,\quad
R_{\m\n} \ra R^\3_{mn},\quad R_{\m\n\r\s} \ra R^\3_{mnrs} 
\eea
and use the following relation
\bea
R^\3_{mnrs}&=&(R^\3_{mr}-\fr14R^\3g^\3_{mr})g^\3_{ns}
-(R^\3_{ms}-\fr14R^\3g^\3_{ms})g^\3_{nr}\nn\\
&&+(R^\3_{ns}-\fr14R^\3g^\3_{ns})g^\3_{ms}
-(R^\3_{nr}-\fr14R^\3g^\3_{nr})g^\3_{ms}
\eea
}  
Let us consider \rf{unsplit} and the following metric redefinition (the tildes have been omitted), 
\bea
g_{\m\n}\ra g_{\m\n}+c_1 \k^2 g_{\m\n}R+c_2 \k^2R_{\m\n};
\eea
the action becomes
\bea
\fr1{\k^2}\int d^4 x \sqrt{-g}\;R &\ra&  \fr1{\k^2}\int d^4 x \sqrt{-g}\;R \nn\\
&&+ \int d^4 x \sqrt{-g}\Big[(c_1+\fr12 c_2)R^2-c_2 R_{\m\n}R^{\m\n}\Big]
\eea
Therefore the counterterms \rf{totalctr} can be absorbed by choosing $c_1,c_2$ appropriately:
{
\bea
c_1+\fr12 c_2={ \fr{1}{240}}\fr{\G(\e)}{(4\pi)^2} 
\quad,\quad -c_2= \fr{41}{120}\fr{\G(\e)}{(4\pi)^2} 
\eea}
These equations yield
{
\bea
c_1={ \fr{7}{40}}\fr{\G(\e)}{(4\pi)^2} \quad,\quad  c_2= -\fr{41}{120}\fr{\G(\e)}{(4\pi)^2} 
\eea	}

\subsection{outline of 1-loop renormalization of $\k$ }

With the counterterms for the one-loop two-point amplitudes determined, we are now in a position to tackle renormalization of the coupling constant. The counterterms in \rf{totalctr} also contribute as the counterterms to one-loop three-point amplitudes shown in Fig.\ref{coupling}. One should compute the one-loop three-point divergence and see whether the aforementioned counterterms are sufficient to remove it. (There are of course ghost loop diagrams to take into account as well; these diagrams will be technically demanding even with help of a computer code.) Unless there is unexpected cancellation, chances are that they will not be sufficient. Then it would be necessary to renormalize the three-point vertex coming from the Einstein-Hilbert action itself; this will be an indication of a need to renormalize the Newton's constant.

\begin{figure}[tbp]
\centering 
\includegraphics[width=.7\textwidth,trim=0 550 -30 0,clip]{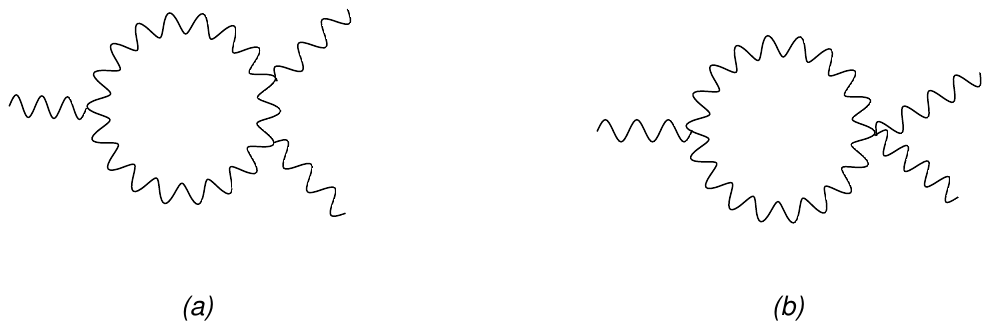}
\caption{3-point graviton amplitudes}
\la{coupling}
\end{figure}

%%%%%%%%%%%%%%%%%%%%%%%%%%%%%%%%%%%%%
%%%%%%%%%%%%%%%%%%%%%%%%%%%%%%%%%%%%%
\section{Conclusion}
%%%%%%%%%%%%%%%%%%%%%%%%%%%%%%%%%%%%%
%%%%%%%%%%%%%%%%%%%%%%%%%%%%%%%%%%%%%

A ``holographic quantization" of the ADM formulation of 4D Einstein gravity was proposed in \cite{Park:2014tia,Park:2014qoa,Park:2014noa} in which a precise identification of the hypersurface of the dynamic degrees of freedom was made with help from the ADM formulation of a non-conventional 3+1 splitting. Carrying out renormalization entirely within a 3D description was proposed therein. In this work we have established, through a 4D covariant approach, renormalizability of the S-matrix based on reduction of the 1PI effective action to 3D. Once only the physical external states are considered, all the fields including the Riemann tensor are reduced to 3D.
With the 4D Riemann tensor reduced to 3D, the effective action becomes expressible in terms of the 3D Ricci tensor and metric, which then leads to renormalizability through a metric field redefinition. Doing things in the entirely 3D way as proposed in \cite{Park:2014tia,Park:2014noa} and doing things 4D covariant way up to a certain point as in the present work should be viewed just as two different renormalization procedures. Presumably they will yield the same results when it comes to physical quantities.

Not surprisingly, 4D covariance has played an important role in establishing renormalizability in the present context. We have addressed two different covariance issues.
The first covariance issue was due to the fact that one considers expansion around a fixed background, and the covariance could be restored by applying the BFM in a refined manner. 
The key was to realize that the following two shift procedures are inequivalent. The first procedure is: shift $g_{\m\n}=\eta_{\m\n}+h_{\m\n}$ and then expand the action. Subsequently make another shift by $h_{\m\n}\ra \vf_{{}_B \m\n}+h_{\m\n}$. This is intrinsically a non-covariant approach and was how the computations were done in \cite{Park:2014noa}. The second approach - the proper one for covariance - is to expand $g_{\m\n}=\eta_{\m\n}+\vf_{{}_B \m\n}+h_{\m\n}$ and treat $\eta_{\m\n}+\vf_{{}_B \m\n}$ as one-piece. Then obtain \rf{gravcubcov} but now by viewing the background metric as $\eta_{\m\n}+\vf_{{}_B \m\n}$. With this, the counterterms have indeed resulted in covariant forms.

The second covariance issue was more obvious: the use of 3D language to describe 4D physics creates tension with 4D covariance.
The crucial step for establishing 4D covariance is to impose the lapse and shift constraint not directly on the starting Lagrangian but rather as the physical state conditions once the off-shell 1PI effective action is obtained. Then the action in the path integral takes the covariant form. The reduction to 3D has taken place at the level of the 1PI action once the external legs are physical states, thus three-dimensional. In other words, the integration is carried out in the full 4D sense; it is the reduction procedure associated with the amplitudes with physical external states that has brought the reduction to 3D.

As a matter of fact, one may further push a question along this line. The question is basically that of gauge-choice independence. (A further discussion can be found in \cite{Park:2015xoa}.) The reduction to 3D seems to depend on our choice of the gauge. Would it be possible to see the reduction with a different choice of gauge? If possible, such a gauge-choice could make things even more covariant. We believe that this question is largely a question on the naturalness of the ADM formalism. As one can see from the review in section 2.1, the reduction of the classical solution to 3D can be established simply by solving the field equations without employing the nonlinear de Donder gauge. The ADM formalism is absolutely natural if one considers a globally hyperbolic spacetime, and many of the physically interesting spacetimes are indeed globally hyperbolic. One may still ask whether it would be possible - for the purpose of making things more covariant - not to solve the constraints explicitly but rather to impose them as constraints throughout. We believe that such an approach will, at best, encounter the type of difficulties that we present in the section ``independent path integral approach" in \cite{Park:2014noa} or perhaps something even worse.

One of the future directions is to explicitly carry out the coupling constant renormalization.
(It would be more desirable to employ more ingredients from the BRST quantization.) Another direction is to extend the analyses of \cite{Park:2014tia} \cite{Park:2014noa} and of the present work to the case of a black hole spacetime.\footnote{This extension has been carried out and recently appeared in \cite{Park:2015xoa}.} We will report on this in the near future.

\vspace{.3in}

\ni {\bf Acknowledgments}

\ni The author thanks K. Lee for hosting the author's visit to KIAS, Seoul. 
The author also thanks S. Sin for his hospitality during the author's visit to Hanyang University, Seoul.

\newpage
\appendix

\newpage
\appendix

\renewcommand{\theequation}{A.\arabic{equation}}
 \setcounter{equation}{0}
\section{Conventions}
The signature is mostly plus: 
%%%
\bea
\eta_{\m\n}=(-,+,+,+)
\eea
%%%
All the Greek indices are four-dimensional
%%%
\bea
\a,\b,\g,...,\m,\n,\r...=0,1,2,3
\eea
%%%
and all the Latin indices are three-dimensional
%%%
\bea
a,b,c,...,m,n,r...=0,1,2
\eea
%%%
The following shorthand notations were used:
%%%
\bea
\f\equiv \eta_{\m\n}\f^{\m\n}\quad,\quad \f^\m\equiv \pa_\k\f^{\k\m}
\eea
%%%
 The 4D graviton and ghost propagators are given by
%%% 
\bea 
<\f_{\m\n}(x_1)\f_{\r\s}(x_2)>&=& P_{\m\n\r\s} \int \fr{d^4k}{(2\pi)^4}\fr{e^{ik\cdot (x_1-x_2)}}{i k^2} \nn\\
<C_\m(x_1)\Cb_\n(x_2)>&=&\eta_{\m\n}\int
 \fr{d^4k}{(2\pi)^4}\;\fr{1}{ik^2}e^{ik\cdot(x_1-x_2)}
\eea
%%%
where, for the traceless propagator,
%%%
\bea
P_{\m\n\r\s}\equiv \fr12(\eta_{\m\r}\eta_{\n\s}+\eta_{\m\s}\eta_{\n\r}- \fr12\eta_{\m\n}\eta_{\r\s})
\eea
%%%

%%%%%%%%%%%%%%%%%%%%%%%%%%%%%%%%%%%%%%%%%%%%
%%%%%%%%%%%%%%%%%%%%%%%%%%%%%%%%%%%%%%%%%%%%
\section{Pathology associated with the traceful propagator}
%%%%%%%%%%%%%%%%%%%%%%%%%%%%%%%%%%%%%%%%%%%%
%%%%%%%%%%%%%%%%%%%%%%%%%%%%%%%%%%%%%%%%%%%%

In the main body it was shown that the counter terms turn out to be covariant once the traceless propagator is used.
In contrast, use of the traceful propagator yields results that cannot be re-written in covariant forms. 
The diagonal kinetic term yields
\bea
-\fr12 < (\cL_{V_I}+\cL_{V_{II}})^2 > 
&=&-\fr12 \fr{\G(\ve)}{(4\pi)^2}\Big[-\fr{7}{12}R^2+\fr{11}{6} R_{\a\b}R^{\a\b}
-\fr7{12}(2R^2 \pa^2\vf+(\pa^2\vf)^2)
\Big]\nn\\
\eea
The cross term  result is
\bea
-\fr12 2< (\cL_{V_I}+\cL_{V_{II}})\cL_{V_{III}} >
&= & -\fr12\fr{\G(\ve)}{(4\pi)^2}\Big[ \Big(4 R^{\a\b}\pa^2 \vf_{\a\b}+\fr53 R\,\pa^\a\pa^\b \vf_{\a\b} -\fr83 R\,\pa^2 \vf  \Big) \Big] \nn\\
\eea
Neither of these can be re-expressed in a covaraint form.
Each term should come out to be covariant if there were no pathology: the pathology must be caused by the trace piece of the fluctuation metric.

\newpage
%%%%%%%%%%%%%%%%%%%%%%%%%%%%%%%%%%%%%%%%%%%%%%%%%%%%%%%%%%%%%%%%


\begin{thebibliography}{99}


%\cite{Stelle:1976gc}
\bibitem{Stelle:1976gc}
  K.~S.~Stelle,
  ``Renormalization of Higher Derivative Quantum Gravity,''
  Phys.\ Rev.\ D {\bf 16} (1977) 953.
  %%CITATION = PHRVA,D16,953;%%
  %1093 citations counted in INSPIRE as of 24 juin 2015



%\cite{Antoniadis:1986tu}
\bibitem{Antoniadis:1986tu} 
  I.~Antoniadis and E.~T.~Tomboulis,
  ``Gauge Invariance and Unitarity in Higher Derivative Quantum Gravity,''
  Phys.\ Rev.\ D {\bf 33}, 2756 (1986).
  %%CITATION = PHRVA,D33,2756;%%
  %102 citations counted in INSPIRE as of 04 Dec 2014





\bibitem{Weinberg3} 
S. Weinberg, in ``General Relativity, an Einstein Centenary Survey,"
edited by S. Hawking and W. Israel, Cambridge (1979)



%\cite{Reuter:1996cp}
\bibitem{Reuter:1996cp} 
  M.~Reuter,
  ``Nonperturbative evolution equation for quantum gravity,''
  Phys.\ Rev.\ D {\bf 57}, 971 (1998)
  [hep-th/9605030].
  %%CITATION = HEP-TH/9605030;%%
  %437 citations counted in INSPIRE as of 01 mar 2015




%\cite{Odintsov:1990qq}
\bibitem{Odintsov:1990qq} 
S.~D.~Odintsov,
``Vilkovisky effective action in quantum gravity with matter,''
Theor.\ Math.\ Phys.\  {\bf 82}, 45 (1990)
[Teor.\ Mat.\ Fiz.\  {\bf 82}, 66 (1990)].
doi:10.1007/BF01028251; S.~D.~Odintsov,
  ``Does the Vilkovisky-De Witt effective action in quantum gravity depend on the configuration space metric?,''
  Phys.\ Lett.\ B {\bf 262}, 394 (1991).
  doi:10.1016/0370-2693(91)90611-S


%\cite{Barvinsky:1993zg}
\bibitem{Barvinsky:1993zg}
A.~O.~Barvinsky, A.~Y.~Kamenshchik and I.~P.~Karmazin,
``The Renormalization group for nonrenormalizable theories: Einstein gravity with a scalar field,''
Phys.\ Rev.\ D {\bf 48} (1993) 3677
doi:10.1103/PhysRevD.48.3677
[gr-qc/9302007].
%%CITATION = doi:10.1103/PhysRevD.48.3677;%%
%58 citations counted in INSPIRE as of 10 Oct 2016




%\cite{VanNieuwenhuizen:1981ae}
\bibitem{VanNieuwenhuizen:1981ae} 
  P.~Van Nieuwenhuizen,
  ``Supergravity,''
  Phys.\ Rept.\  {\bf 68}, 189 (1981).
  %%CITATION = PRPLC,68,189;%%
  %1133 citations counted in INSPIRE as of 24 juin 2015


%\cite{Bern:2011qn}
\bibitem{Bern:2011qn}
  Z.~Bern, J.~J.~Carrasco, L.~J.~Dixon, H.~Johansson and R.~Roiban,
  ``Amplitudes and Ultraviolet Behavior of N = 8 Supergravity,''
  Fortsch.\ Phys.\  {\bf 59} (2011) 561
  [arXiv:1103.1848 [hep-th]].
  %%CITATION = ARXIV:1103.1848;%%
  %45 citations counted in INSPIRE as of 24 juin 2015




%\cite{Ashtekar:1986yd}
\bibitem{Ashtekar:1986yd}
  A.~Ashtekar,
  ``New Variables for Classical and Quantum Gravity,''
  Phys.\ Rev.\ Lett.\  {\bf 57}, 2244 (1986).
  %%CITATION = PRLTA,57,2244;%%
  %912 citations counted in INSPIRE as of 11 Apr 2014



%\cite{Thiemann:2007zz}
\bibitem{Thiemann:2007zz}
  T.~Thiemann,
  ``Modern canonical quantum general relativity,''
  Cambridge, UK: Cambridge Univ. Pr. (2007) 819 p
  [gr-qc/0110034].
  %%CITATION = GR-QC/0110034;%%
  %486 citations counted in INSPIRE as of 11 Apr 2014



%\cite{Ambjorn:2012jv}
\bibitem{Ambjorn:2012jv} 
J.~Ambjorn, A.~Goerlich, J.~Jurkiewicz and R.~Loll,
``Nonperturbative Quantum Gravity,''
Phys.\ Rept.\  {\bf 519}, 127 (2012)
doi:10.1016/j.physrep.2012.03.007
[arXiv:1203.3591 [hep-th]].
%%CITATION = doi:10.1016/j.physrep.2012.03.007;%%
%102 citations counted in INSPIRE as of 14 May 2016



%\cite{Calcagni:2012hb}
\bibitem{Calcagni:2012hb} 
  G.~Calcagni,
  ``Introduction to multifractional spacetimes,''
  AIP Conf.\ Proc.\  {\bf 1483}, 31 (2012)
  doi:10.1063/1.4756961
  [arXiv:1209.1110 [hep-th]].
  %%CITATION = doi:10.1063/1.4756961;%%
  %12 citations counted in INSPIRE as of 11 May 2017


%\cite{Donoghue:2015hwa}
\bibitem{Donoghue:2015hwa} 
J.~F.~Donoghue and B.~R.~Holstein,
``Low Energy Theorems of Quantum Gravity from Effective Field Theory,''
J.\ Phys.\ G {\bf 42}, no. 10, 103102 (2015)
doi:10.1088/0954-3899/42/10/103102
[arXiv:1506.00946 [gr-qc]].
%%CITATION = doi:10.1088/0954-3899/42/10/103102;%%
%7 citations counted in INSPIRE as of 17 Jun 2016




%\cite{York:1972sj}
\bibitem{York:1972sj} 
  J.~W.~York, Jr.,
  ``Role of conformal three geometry in the dynamics of gravitation,''
  Phys.\ Rev.\ Lett.\  {\bf 28}, 1082 (1972).
  %%CITATION = PRLTA,28,1082;%%
  %532 citations counted in INSPIRE as of 12 Feb 2015




%\cite{Moncrief:1989dx}
\bibitem{Moncrief:1989dx} 
  V.~Moncrief,
  ``Reduction of the Einstein equations in (2+1)-dimensions to a Hamiltonian system over Teichmuller space,''
  J.\ Math.\ Phys.\  {\bf 30}, 2907 (1989).
  %%CITATION = JMAPA,30,2907;%%
  %128 citations counted in INSPIRE as of 04 Sep 2014


%\cite{Fischer:1996qg}
\bibitem{Fischer:1996qg} 
  A.~E.~Fischer and V.~Moncrief,
  ``Hamiltonian reduction of Einstein's equations of general relativity,''
  Nucl.\ Phys.\ Proc.\ Suppl.\  {\bf 57}, 142 (1997).
  %%CITATION = NUPHZ,57,142;%%
  %10 citations counted in INSPIRE as of 04 Sep 2014



%\cite{Gay-Balmaz:2014ena}
\bibitem{Gay-Balmaz:2014ena} 
  F.~Gay-Balmaz and T.~S.~Ratiu,
  ``A new Lagrangian dynamic reduction in field theory,''
  Annales Inst.\ Fourier {\bf 16}, 1125 (2010)
  [arXiv:1407.0263 [math-ph]].
  %%CITATION = ARXIV:1407.0263;%%
  %2 citations counted in INSPIRE as of 17 Jan 2015





%\cite{Sato:2002kv}
\bibitem{Sato:2002kv}
  M.~Sato and A.~Tsuchiya,
  ``Born-Infeld action from supergravity,''
  Prog.\ Theor.\ Phys.\  {\bf 109}, 687 (2003)
  [hep-th/0211074].
  %%CITATION = HEP-TH/0211074;%%
  %24 citations counted in INSPIRE as of 24 Feb 2014




%\cite{Park:2013vpa}
\bibitem{Park:2013vpa} 
  I.~Y.~Park,
  ``Dimensional reduction to hypersurface of foliation,''
  Fortsch.\ Phys.\  {\bf 62}, 966 (2014)
  [arXiv:1310.2507 [hep-th]].
  %%CITATION = ARXIV:1310.2507;%%
  %4 citations counted in INSPIRE as of 23 Feb 2015




%\cite{Park:2014tia}
\bibitem{Park:2014tia} 
  I.~Y.~Park,
  ``Hypersurface foliation approach to renormalization of ADM formulation of gravity,''
  Eur.\ Phys.\ J.\ C {\bf 75}, no. 9, 459 (2015)
  doi:10.1140/epjc/s10052-015-3660-x
  [arXiv:1404.5066 [hep-th]].
  %%CITATION = doi:10.1140/epjc/s10052-015-3660-x;%%
  %6 citations counted in INSPIRE as of 08 Feb 2016




%\cite{Hatefi:2012bp}
\bibitem{Hatefi:2012bp} 
E.~Hatefi, A.~J.~Nurmagambetov and I.~Y.~Park,
``ADM reduction of IIB on $\mathcal{H}^{p,q}$ to dS braneworld,''
JHEP {\bf 1304}, 170 (2013)
doi:10.1007/JHEP04(2013)170
[arXiv:1210.3825 [hep-th]].
%%CITATION = doi:10.1007/JHEP04(2013)170;%%
%27 citations counted in INSPIRE as of 02 Sep 2016




%\cite{Higuchi:1991tk}
\bibitem{Higuchi:1991tk} 
  A.~Higuchi,
  ``Quantum linearization instabilities of de Sitter space-time. 1,''
  Class.\ Quant.\ Grav.\  {\bf 8}, 1961 (1991).
  doi:10.1088/0264-9381/8/11/009
  %%CITATION = doi:10.1088/0264-9381/8/11/009;%%
  %28 citations counted in INSPIRE as of 11 May 2017







%\cite{Park:2014qoa}
\bibitem{Park:2014qoa} 
  I.~Y.~Park,
  ``Quantization of gravity through hypersurface foliation,''
  arXiv:1406.0753 [gr-qc].
  %%CITATION = ARXIV:1406.0753;%%
  %2 citations counted in INSPIRE as of 18 Jan 2015

%\cite{Park:2015qxa}
\bibitem{Park:2015qxa} 
I.~Y.~Park,
``Foliation, jet bundle and quantization of Einstein gravity,''
Front.\ in Phys.\  {\bf 4}, 25 (2016)
doi:10.3389/fphy.2016.00025
[arXiv:1503.02015 [hep-th]].
%%CITATION = doi:10.3389/fphy.2016.00025;%%
%5 citations counted in INSPIRE as of 29 Jul 2016



%\cite{Park:2014noa}
\bibitem{Park:2014noa} 
I.~Y.~Park,
``Lagrangian constraints and renormalization of 4D gravity,''
JHEP {\bf 1504}, 053 (2015)
[arXiv:1412.1528 [hep-th]].
%%CITATION = ARXIV:1412.1528;%%
%1 citations counted in INSPIRE as of 24 Apr 2015



%\cite{Mazur:1989by}
\bibitem{Mazur:1989by} 
  P.~O.~Mazur and E.~Mottola,
  ``The Gravitational Measure, Solution of the Conformal Factor Problem and Stability of the Ground State of Quantum Gravity,''
  Nucl.\ Phys.\ B {\bf 341}, 187 (1990).
  %%CITATION = NUPHA,B341,187;%%
  %108 citations counted in INSPIRE as of 04 juin 2015



\bibitem{Buchbinder}
I. L. Buchbinder, S. D. Odintsov and I. L. Shapiro, ``Effective action in quantum gravity," IOP Publishing Ltd (1992)


%\cite{Gibbons:1978ac}
\bibitem{Gibbons:1978ac} 
  G.~W.~Gibbons, S.~W.~Hawking and M.~J.~Perry,
  ``Path Integrals and the Indefiniteness of the Gravitational Action,''
  Nucl.\ Phys.\ B {\bf 138}, 141 (1978).
  %%CITATION = NUPHA,B138,141;%%
  %405 citations counted in INSPIRE as of 04 juin 2015




%\cite{'tHooft:1973us}
\bibitem{'tHooft:1973us} 
G.~'t Hooft,
``An algorithm for the poles at dimension four in the dimensional regularization procedure,''
Nucl.\ Phys.\ B {\bf 62}, 444 (1973).
%%CITATION = NUPHA,B62,444;%%
%460 citations counted in INSPIRE as of 07 Mar 2015


%\cite{Park:2015xoa}
\bibitem{Park:2015xoa} 
I.~Y.~Park,
``Holographic quantization of gravity in a black hole background,''
J.\ Math.\ Phys.\  {\bf 57}, no. 2, 022305 (2016)
doi:10.1063/1.4942101
[arXiv:1508.03874 [hep-th]].
%%CITATION = doi:10.1063/1.4942101;%%
%1 citations counted in INSPIRE as of 14 Mar 2016





%\cite{Kuchar:1970mu}
\bibitem{Kuchar:1970mu} 
  K.~Kuchar,
  ``Ground state functional of the linearized gravitational field,''
  J.\ Math.\ Phys.\  {\bf 11}, 3322 (1970).
  %%CITATION = JMAPA,11,3322;%%
  %52 citations counted in INSPIRE as of 04 juin 2015


\bibitem{Ortin}

T. Ortin, ``Gravity and strings," Cambridge University Press (2004)



%\cite{Antoniadis:1995fc}
\bibitem{Antoniadis:1995fc} 
  I.~Antoniadis, J.~Iliopoulos and T.~N.~Tomaras,
  ``One loop effective action around de Sitter space,''
  Nucl.\ Phys.\ B {\bf 462}, 437 (1996)
  doi:10.1016/0550-3213(95)00633-8
  [hep-th/9510112].
  %%CITATION = doi:10.1016/0550-3213(95)00633-8;%%
  %32 citations counted in INSPIRE as of 21 Apr 2017



%\cite{'tHooft:1974bx}
\bibitem{'tHooft:1974bx}
  G.~'t Hooft and M.~J.~G.~Veltman,
  ``One loop divergencies in the theory of gravitation,''
  Annales Poincare Phys.\ Theor.\ A {\bf 20}, 69 (1974).
  %%CITATION = AHPAA,A20,69;%%
  %756 citations counted in INSPIRE as of 10 Apr 2014





%\cite{Deser:1974cz}
\bibitem{Deser:1974cz}
  S.~Deser and P.~van Nieuwenhuizen,
  ``One Loop Divergences of Quantized Einstein-Maxwell Fields,''
  Phys.\ Rev.\ D {\bf 10}, 401 (1974).
  %%CITATION = PHRVA,D10,401;%%
  %347 citations counted in INSPIRE as of 10 Apr 2014




%\cite{Goroff:1985th}
\bibitem{Goroff:1985th}
  M.~H.~Goroff and A.~Sagnotti,
  ``The Ultraviolet Behavior of Einstein Gravity,''
  Nucl.\ Phys.\ B {\bf 266}, 709 (1986).
  %%CITATION = NUPHA,B266,709;%%
  %293 citations counted in INSPIRE as of 12 Mar 2014










%\cite{Capper:1973pv}
\bibitem{Capper:1973pv} 
  D.~M.~Capper, G.~Leibbrandt and M.~Ramon Medrano,
  ``Calculation of the graviton selfenergy using dimensional regularization,''
  Phys.\ Rev.\ D {\bf 8}, 4320 (1973).
  %%CITATION = PHRVA,D8,4320;%%
  %115 citations counted in INSPIRE as of 24 Mar 2015



%\cite{Arnowitt:1962hi}
\bibitem{Arnowitt:1962hi} 
  R.~L.~Arnowitt, S.~Deser and C.~W.~Misner,
  ``The Dynamics of general relativity,''
  Gen.\ Rel.\ Grav.\  {\bf 40}, 1997 (2008)
  [gr-qc/0405109].
  %%CITATION = GR-QC/0405109;%%
  %638 citations counted in INSPIRE as of 10 Apr 2014




%\cite{Poisson}
\bibitem{Poisson} 
E. Poisson, ``A relativists' toolkit", Cambridge (2004)

%\cite{Park:2016zgt}
\bibitem{Park:2016zgt} 
  I.~Y.~Park,
  ``One-loop renormalization of a gravity-scalar system,'' to appear in EPJC,
  arXiv:1606.08384 [hep-th].
  %%CITATION = ARXIV:1606.08384;%%


\bibitem{Weinberg2}

S. Weinberg, ``The quantum theory of fields", vol II, Cambridge university press (1995)


\bibitem{Zinn-Justin}

J. Zinn-Justin, ``Quantum field theory and critical phenomena," 3rd Ed., Oxford university press (1996) 


\bibitem{Sterman}

G. Sterman, ``An introduction to quantum field theory," Cambridge university press (1993)


















\end{thebibliography}
\end{document}